\UseRawInputEncoding
\documentclass[%
 reprint,
 showkeys,
 amsmath,amssymb,
 aps,
twocolumn,
superscriptaddress,
altaffilletter
]{revtex4-2}

\usepackage[%
    colorlinks=true,
    pdfborder={0 0 0},
    linkcolor=blue,
    allcolors=blue
]{hyperref}
\usepackage{nameref} 
\makeatletter
\newcommand{\labelname}[1]{
  \def\@currentlabelname{#1}}%
\makeatother
\usepackage{graphicx}
\usepackage{dcolumn}
\usepackage{bm}
\usepackage[dvipsnames]{xcolor}
\usepackage{stmaryrd}
\usepackage{float}
\usepackage{booktabs}

\newcommand{\beginsupplement}{%
        \setcounter{table}{0}
        \renewcommand{\thetable}{S\arabic{table}}%
        \setcounter{figure}{0}
        \renewcommand{\thefigure}{S\arabic{figure}}%
     }
\newcommand{\bra}[1]{\langle #1 \vert}
\newcommand{\ket}[1]{\vert #1 \rangle}

\newcommand{\gate}[1]{\mathsf {#1} }

\begin{document}

\preprint{APS/123-QED}

\title{Calibrated decoders for experimental quantum error correction}

\author{Edward H. Chen}
\altaffiliation{%
 Equally contributing spokesperson
}%
\email{ehchen@ibm.com}
\affiliation{%
 Almaden Research Center, San Jose, CA 95120, USA
}%
\author{Theodore J. Yoder}%
\altaffiliation{%
 Equally contributing spokesperson
}%
\email{ted.yoder@ibm.com}
\affiliation{%
 T.J. Watson Research Center, Yorktown Heights, NY 10598, USA
}%

\author{Youngseok Kim}%
\author{Neereja Sundaresan}%

\author{Srikanth Srinivasan}%

\author{Muyuan Li}%
\author{Antonio D. C\'orcoles}%

\author{Andrew W. Cross}%

\author{Maika Takita}%
\affiliation{%
 T.J. Watson Research Center, Yorktown Heights, NY 10598, USA
}%

\collaboration{IBM Quantum}

\date{\today}

\begin{abstract}
Arbitrarily long quantum computations require quantum memories that can be repeatedly measured without being corrupted. 
Here, we preserve the state of a quantum memory, notably with the additional use of flagged error events. 
All error events were extracted using fast, mid-circuit measurements and resets of the physical qubits. 
Among the error decoders we considered, we introduce a perfect matching decoder that was calibrated from measurements containing up to size-4 correlated events.
To compare the decoders, we used a \textit{partial} post-selection scheme shown to retain ten times more data than \textit{full} post-selection.
We observed logical errors per round of $2.2\pm0.1\times10^{-2}$ (decoded \textit{without} post-selection) and $5.1\pm0.7\times10^{-4}$ (\textit{full} post-selection), which was less than the physical measurement error of $7\times10^{-3}$ and therefore surpasses a pseudo-threshold for repeated logical measurements.
\end{abstract}

\maketitle

\section{\label{sec:intro}Introduction}

Preparing and preserving logical quantum states is necessary for performing long quantum computations~\cite{gouzienFactoring2048bitRSA2021}. Because noise inevitably corrupts the underlying physical qubits, quantum error correction (QEC) codes have been designed to detect and recover from errors~\cite{aharonovFaultTolerantQuantumComputation2008, knillThresholdAccuracyQuantum1996a, gottesmanStabilizerCodesQuantum1997, shorSchemeReducingDecoherence1995,knillTheoryQuantumErrorcorrecting1997}. Significant efforts are currently focused on demonstrating capabilities that will be necessary for implementing practical QEC. An optimal choice of a code varies depending on the device and its noise properties~\cite{iyerSmallQuantumComputer2018}. Notable experimental implementations include NMR~\cite{moussaDemonstrationSufficientControl2011,zhangExperimentalQuantumError2011}, ion traps~\cite{niggQuantumComputationsTopologically2014a,linkeFaulttolerantQuantumError2017, eganFaulttolerantControlErrorcorrected2021b, ryan-andersonRealizationRealtimeFaulttolerant2021}, donors~\cite{waldherrQuantumErrorCorrection2014,abobeihFaulttolerantOperationLogical2021,hillExchangebasedSurfacecodeQuantum2021}, quantum dots~\cite{xueComputingSpinQubits2021,andrewsQuantifyingErrorLeakage2019}, and superconducting qubits~\cite{reedRealizationThreequbitQuantum2012, kellyStatePreservationRepetitive2015, corcolesDemonstrationQuantumError2015c, risteDetectingBitflipErrors2015, takitaExperimentalDemonstrationFaultTolerant2017c}. Recent developments of high-fidelity mid-circuit measurements and resets of superconducting qubits have enabled the preparation and repeated stabilization of logical states~\cite{andersenRepeatedQuantumError2020a, marquesLogicalqubitOperationsErrordetecting2021,googlequantumaiExponentialSuppressionBit2021}; demonstrations of such quantum memories with enhanced lifetimes have been limited by, among other reasons, a combination of gate and measurement cross-talk.

One way to mitigate cross-talk~\cite{takitaDemonstrationWeightFourParity2016} is to reduce the lattice connectivity~\cite{hertzbergLaserannealingJosephsonJunctions2021,ibmquantumHighfidelitySuperconductingQuantum2020}. Consequently, fault-tolerant operations require intermediary qubits; such qubits can be used to \textit{flag} high-weight errors originating from low-weight errors~\cite{chaoQuantumErrorCorrection2018, chamberlandFlagFaulttolerantError2018}. In certain QEC codes and lattice geometries, flag qubits supply the information needed to extend the effective distance of a QEC code up to its intended distance, and thus enable maximal efficiency at detecting and correcting errors~\cite{chamberlandTopologicalSubsystemCodes2020}.

\begin{figure}[H]
\includegraphics[width=0.9\columnwidth]{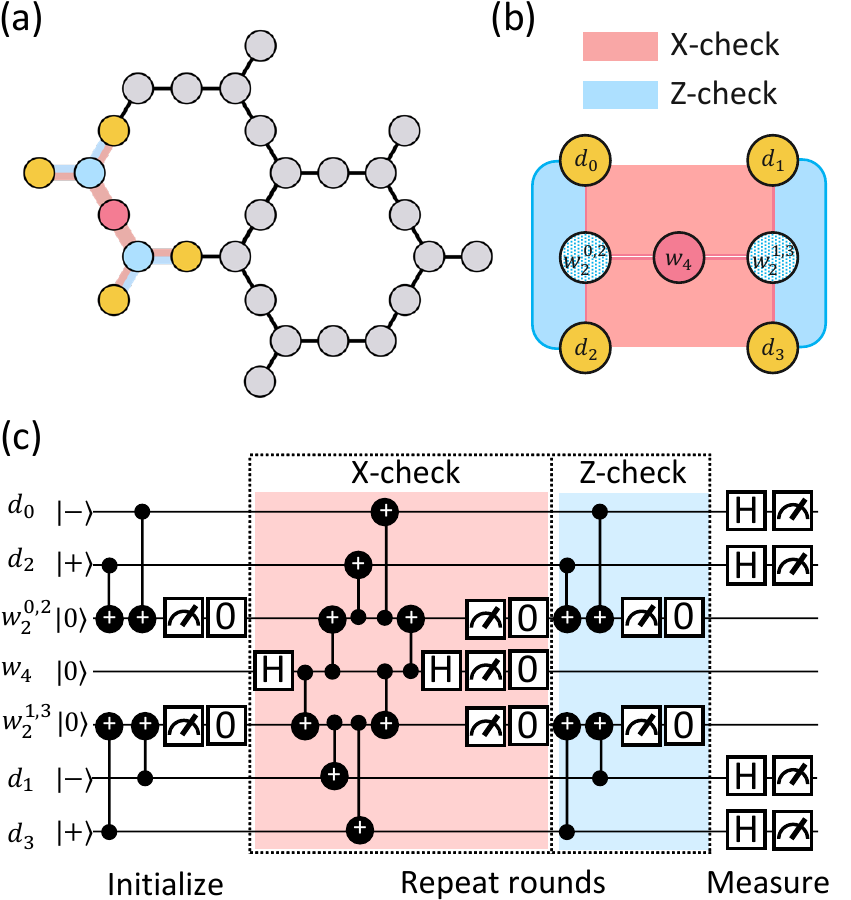}
\caption{(Color)
(a) Experiments were performed on \textit{ibmq\_kolkata}, which had 27 qubits connected in a heavy hexagon (HH) topology. The 7 qubits used for the $\llbracket4,1,2\rrbracket$ code are colored yellow, blue and pink. 
(b) The code layout indicates a single weight-4, X stabilizer (pink), and two weight-2, Z stabilizers (blue) on the four data qubits (yellow) labelled $d_i$ for integers `i' from 0 to 3. For the weight-2 stabilizers, superscripts `0,2' (`1,3') indicates the data qubits on which they operate. The reduced connectivity of the graph is addressed by flag qubits (blue) alternating between (i) being used as weight-2 stabilizers, and (ii) as intermediary qubits used to detect errors on the center, syndrome qubit (pink).
(c) Circuit diagram for the code layout in (b) applied to an initial $|-\rangle_L$ logical state with alternating, repeated X- and Z-check stabilizer measurements, together comprising a round, with mid-circuit reset operations (`0') applied between rounds. In this illustration, the final measurement measures the four data qubits in the X-basis.
}
\label{fig:layout}
\end{figure}

We demonstrated repeated error detection and correction of a $\llbracket4,1,2\rrbracket$ error-detecting topological stabilizer code on a heavy-hexagonal (HH) device designed to mitigate the limiting effects of cross-talk using flag qubits. The combination of fast readout with reduced qubit connectivity improved, after post-selecting on instances in which no errors were detected, logical errors per round when compared to the physical measurement error rate. A thorough analysis of this code led us to introduce a \textit{partial} post-selection scheme allowing us to discard ten times less data for comparing matching decoding algorithms. Compared against previously known decoding strategies on the entire data set, we found that a decoder performed best with experimentally-calibrated edge weights that account for the correlations between syndromes. Furthermore, we showed that correlations between five or more syndromes can be eliminated by the application of a ``deflagging'' procedure. The minimal impact of deflagging on logical errors is an encouraging sign that this technique, and its extension to general flag-based codes, is a viable way to process flag outcomes in practice.

\section{\label{sec:theory}Theory}
\begin{figure}[b]
\includegraphics{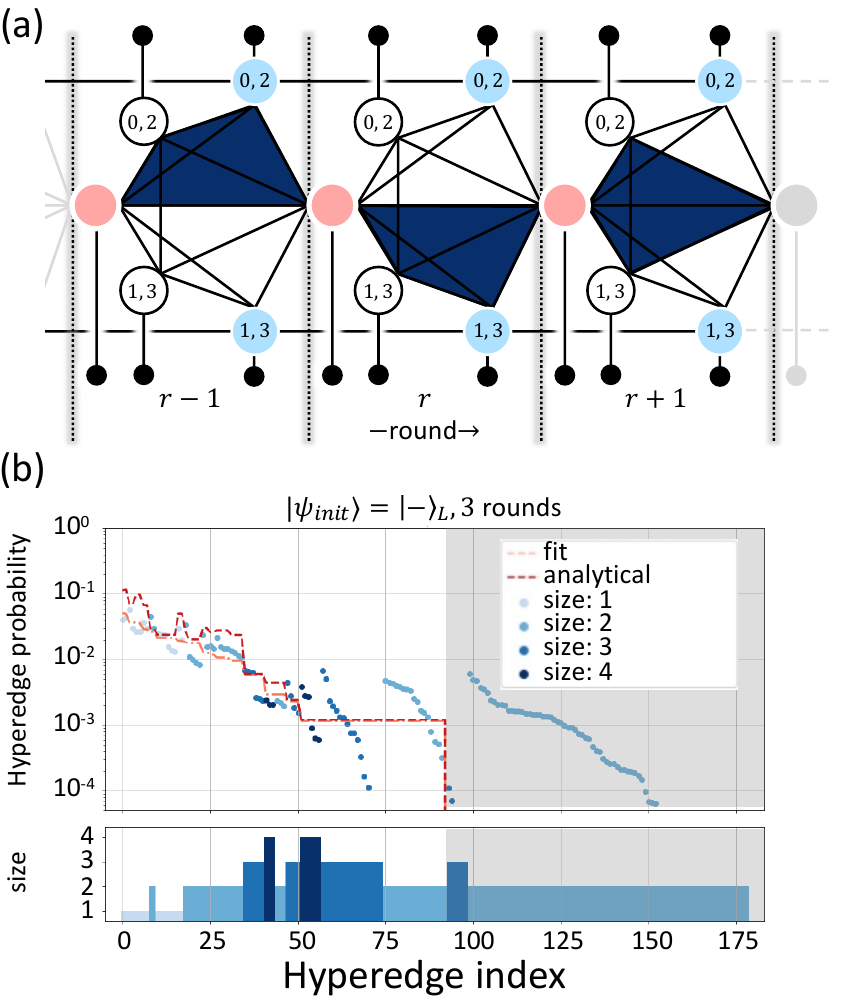}
\caption{(Color)
(a) Decoder graph for the code layout depicted in Fig.~\ref{fig:layout}(b). Syndromes from weight-4 (2) stabilizers are mapped to the pink (blue) nodes, and the weight-2 flag measurements are mapped onto the white nodes. Identical to Fig.~\ref{fig:layout}, `0,2' (`1,3') denotes the left (right) hand side of the code layout. For initial $|-/+\rangle_L$ states stabilized by the circuit in Fig.~\ref{fig:layout}(c), there are three different possible size-4 hyperedges within each round, each highlighted in dark blue across three consecutive rounds. The boundary nodes in black have, by definition, edges with weight `0' connecting them; rendering all boundary nodes to be effectively a single node for the purposes of the decoding process.
(b) Experimentally adjusted correlation probabilities of hyperedges for the circuit in Fig.~\ref{fig:layout}(c) with three rounds of stabilizer measurements. 
The hyperedges are sorted from largest to smallest based on the results from a least-squares fit using a six-parameter noise model. Points with darker colors represent hyperedges of greater sizes, as shown in the lower half of the plot. Hyperedges with indices greater than 93 (shaded gray) had no analytical expression, but were still experimentally adjusted to quantify the impact of computational leakage (Supp.~\nameref{sec:paulitracing}). The result of fitting the six-parameter noise model (fit, pink dash) agreed well with the analytical (red dash) curve generated using noise terms from simultaneous randomized benchmarking (Table~\ref{table:noiseSix}).
}
\label{fig:fig2} 
\end{figure}

Active error-correction involves decoding, using syndrome measurements, the errors that occurred in the circuit so that the proper corrections can be applied. We define error-sensitive events to be linear combinations of syndrome measurement bits that, in an ideal circuit, would be zero. Thus, a non-zero error-sensitive event indicates some error has occurred. For the HH code, there are two types of error-sensitive events (1) the difference of two subsequent measurements of the same stabilizer, and (2) flag qubit measurements.

Error-sensitive events are depicted as nodes in a decoding graph with edges representing errors that are detected by both events at their end points (Fig.~\ref{fig:fig2}(a)). If the probability an edge occurs is $P$, then the edge is given weight $\log((1-P)/P)$. The decoding graph may also have a boundary node, so that an error detected by just one error-sensitive event can be represented as an edge from that event to the boundary node. In practice, there are also errors detected by more than two error-sensitive events that could be represented as \textit{hyper}edges in a more general decoding \textit{hyper}graph.

Given a set of non-zero error-sensitive events, minimum-weight perfect-matching (MWPM) finds paths of edges connecting pairs of those events with minimum total weight, and is a simple and effective decoding algorithm for a topological stabilizer code that only operates on a decoding \textit{graph}~\cite{higgottPyMatchingPythonPackage2021}, as opposed to a decoding \textit{hypergraph}.  While MWPM is computationally efficient, the analogous matching algorithm on a hypergraph is not, which limits the practicality of a decoding hypergraph.

The effectiveness of MWPM depends crucially on edge weights in the decoding graph. We explored three strategies for setting these edge weights: (1) In the \textit{uniform} approach, all edge weights were identical. (2) In the \textit{analytical} approach, edge weights were individually calculated in terms of Pauli error rate parameters $\rho_j$, where the index $j$ indicates one of the six errors being considered: CNOT gates, single-qubit gates, idle locations, initialization, resets, and measurements. The numerical values of the parameters $\rho_j$ can be chosen in several ways as discussed in Sect.~\ref{sec:experiment}. (3) In the \textit{correlation} approach, we analyzed experimental data to determine a set of edge probabilities that are likely to have produced it. 
This approach involved first calculating the probabilities for all hyperedges in the decoding \textit{hypergraph} before determining the edge probabilities used in the decoder \textit{graph}.

A hyperedge in the decoding hypergraph represents any of a number of Pauli faults in the circuit that are indistinguishable from one another because they each lead to the same set of non-zero error-sensitive events. If several faults occur together, the symmetric difference of their hyperedges is denoted $S$, the syndrome, or, in other words, the set of non-zero error-sensitive events that is observed. The probability we observe a particular $S$ is the probability that hyperedges occur in combination to produce $S$. Since this is related to the probability $\alpha_h$ of an individual hyperedge $h$ occurring, we can learn $\alpha_h$ from many observations of $S$.

Realistically, the possible hyperedges are limited in size $|h|$ by locality of the circuits. In the $\llbracket4,1,2\rrbracket$ code, we found that hyperedges are limited to sizes four or less. Finding $\alpha_h$ in practical time begins by considering local clusters and then adjusting local estimates recursively from size-4 hyperedges down to size-1 and -2 (Fig.~\ref{fig:fig2} and \hyperref[sec:CorrelationAnalysis]{Supp.~\nameref{sec:CorrelationAnalysis}}). Only size-1 and -2 edges are required for MWPM, but ignoring larger hyperedges can result in nonphysical, negative size-1 correlations.

Another way we explored decoding strategies was to consider analyzing only a subset of all data. By Pauli tracing (\hyperref[sec:paulitracing]{Supp.~\nameref{sec:paulitracing}}), we classified edges in the decoding graph into three categories depending on whether its inclusion in the minimum-weight matching necessitated: (1) flipping the logical measurement, (2) not flipping the logical measurement, or (3) is ambiguous (Fig.~\ref{fig:decoderEdges}). The ambiguous case occurs specifically for error-detecting codes, like the $\llbracket4,1,2\rrbracket$ code presented here, because some errors result in the decoder having to choose between two equally probable corrections. 

Using these classifications for edges in the decoder graph, we explored three degrees of post-selection. The most conservative approach, using \textit{full} post-selection, involved discarding all results showing any non-zero error-sensitive event; this approach was the only one in which further decoding cannot be done. In the opposite regime, \textit{without} post-selection, all results were kept and any ambiguous edges in the MWPM were treated without flipping the logical measurement; here, logical error rates could have been improved by decoding but was not strictly needed. Finally, the intermediate regime involved a \textit{partial} post-selection scheme whereby results were only discarded if the MWPM algorithm highlighted an ambiguous edge; here, decoding had to be done so that results with ambiguous edges that were highlighted could be discarded.

\begin{figure}[b]
\includegraphics{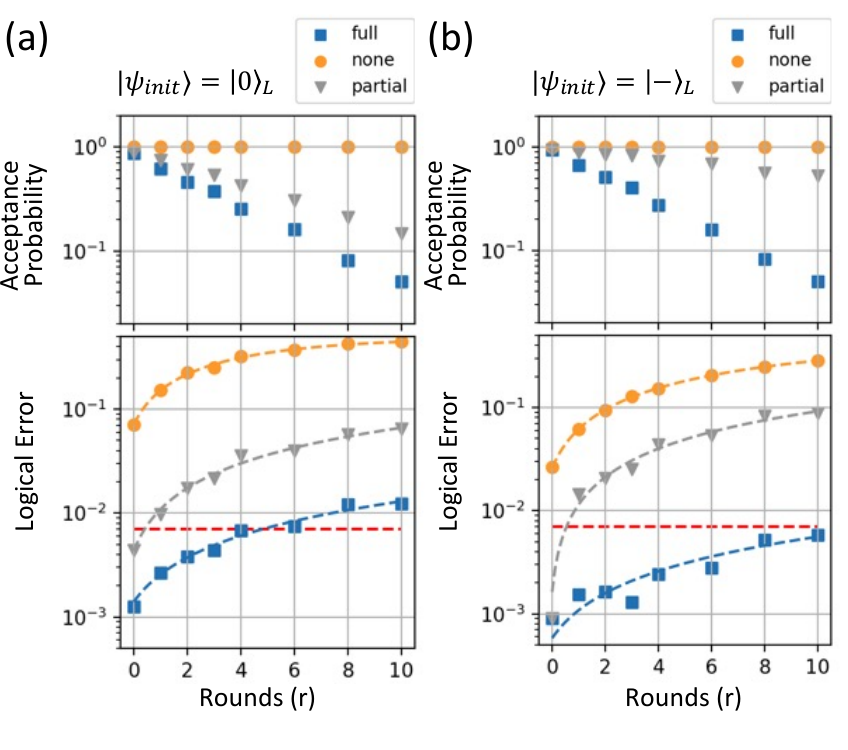}
\caption{(Color)
\label{fig:fig3}
(top) Fraction of total results used for the logical states ((a) $|0\rangle_L$, (b) $|-\rangle_L$) as the number of stabilizer rounds were repeated from 0 to 10 times when \textit{full} (blue squares), \textit{none} (yellow circles), or \textit{partial} (gray triangles) post-selection analysis was used. 
(bottom) The corresponding logical errors versus number of rounds. The dashed red lines indicate the pseudo-threshold as determined by the best (average) physical measurement errors of $7\times10^{-3}$ ($7.7\times10^{-3}$).
}
\end{figure}

\section{\label{sec:experiment}Experimental results}

Fitting the adjusted hyperedge probabilities to analytical expressions produces approximate estimates for the six-noise parameters in the error correcting experiments (Fig.~\ref{fig:fig2}(b)). These noise estimates were found to be in good agreement with benchmarks based on simultaneous randomized benchmarking. Experiments were performed on four logical states ($|-/+\rangle_L$ and $|0/1\rangle_L$) each of which was stabilized up to 10 rounds to extract a logical error per round of stabilizers (Fig.~\ref{fig:fig3}). This logical error varied depending on the analysis method. 

For the \textit{full} post-selection scheme, the logical error for some rounds fell below the best and average physical initialization and measurement errors - a hallmark of being below a so-called pseudo-threshold for fault-tolerant quantum computing. 
Fitting the decay curves resulted in inferred logical errors per round of $6.4\pm1.3\times10^{-4}$ for $|-/+\rangle_L$, and $11\pm1 \times 10^{-4}$ for $|0/1\rangle_L$. 

If \textit{none} of the instances of the experiment were discarded, then the logical error remained consistently above the pseudo-threshold. 
In this analysis without any post-selection and without decoding, we inferred logical errors per round of $40.4\pm0.2 \times10^{-3}$ for $|-/+\rangle_L$, and $102\pm2 \times 10^{-3}$ for $|0/1\rangle_L$.

Recalling that the $\llbracket4,1,2\rrbracket$ is an error detecting code, we used the syndrome outcomes from each stabilizer round to perform a \textit{post-facto} logical correction in software. 
Discarding instances where ambiguous edges were highlighted by the decoder allowed us to apply the \textit{partial}, in contrast with the \textit{full}, post-selection scheme. With this scheme, significantly more instances of the experimental runs remained, resulting in inferred logical errors per round of $10.7\pm0.7\times 10^{-3}$ for $|-/+\rangle_L$, and $6.2\pm0.3\times 10^{-3}$ for $|0/1\rangle_L$.
\onecolumngrid

\begin{figure}[h]
\includegraphics{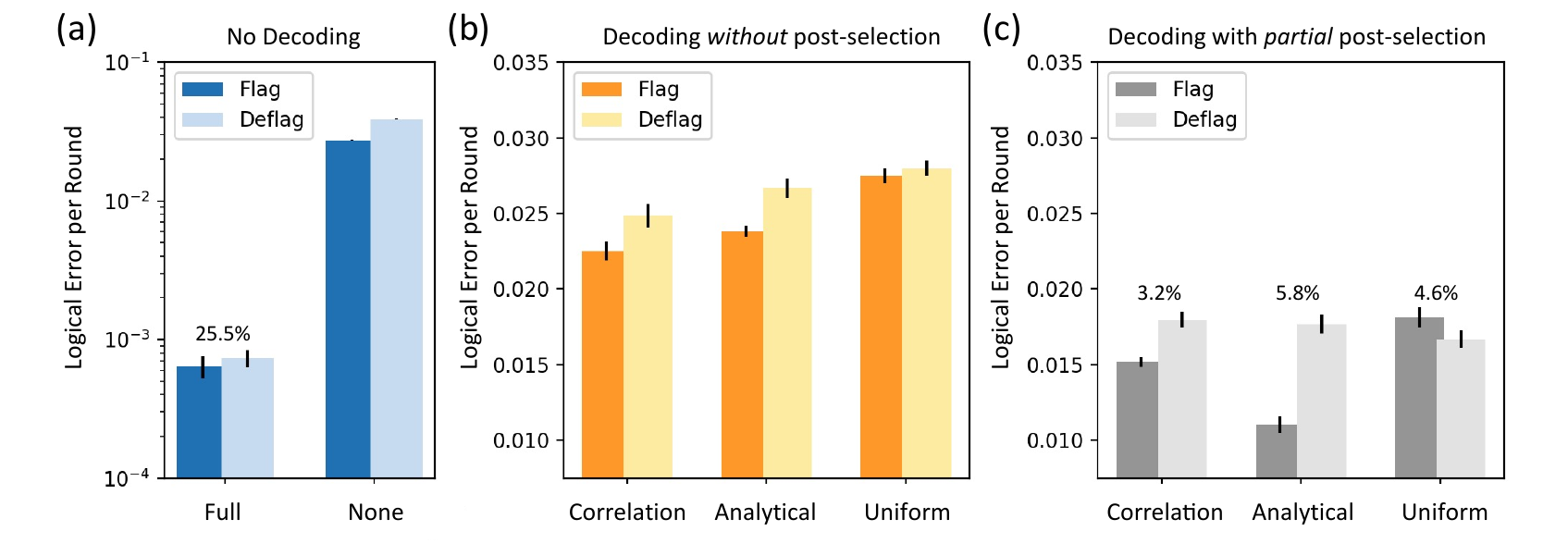}
\caption{(Color)
Logical errors per round initially in $|-/+\rangle_L$ states under various analysis methods with acceptance probability per round labelled above. Results varied depending on whether the flag events were directly used for decoding (\textit{flag}) or indirectly used for decoding using a \textit{deflagging} procedure~(\hyperref[sec:deflagging]{Supp.~\nameref{sec:deflagging}}).
(a) Logical error per round for \textit{full} and \textit{none} post-selection methods. 25.5\% of the counts were rejected with each round for the \textit{full} post-selection scheme.
(b) Comparison between errors using three decoder graphs on data \textit{without} post-selection. 
(c) Comparison between errors using three decoder graphs on data with \textit{partial} post-selection. The approximate percentage of counts rejected for each stabilizer round are indicated above each bar.
}
\label{fig:fig4}
\end{figure}

\twocolumngrid

Within the \textit{none} and \textit{partial} post-selection schemes, we were able to compare the performance of three different instances of decoders~(Fig.~\ref{fig:fig4}). 
The most generic decoder assumes there was no known noise model for the underlying physical system. Such a \textit{uniform} decoder graph, in which every edge of the decoder graph was given equal weights, was expected to perform better than no decoding at all; but, was expected to be worse than any other graph whose edges were informed by some knowledge of the underlying noise. For instance, by selecting a simple, Pauli noise model, analytical expressions for the edge weights were calculated and led to improved logical error rates. Alternatively, if no assumptions were made about the noise, then edge weights were populated by the experimentally calibrated, correlation probabilities described earlier. We found that, as expected, such a correlation decoder graph indeed corrected for logical errors more effectively than the uniform decoding strategy and compared well with the analytical method (Fig.~\ref{fig:fig4}(b)). However, when the \textit{partial} post-selection scheme was used, this trend no longer held since an analytical decoder with noise parameters from simultaneous randomized benchmarking outperformed the correlation analysis (Fig.~\ref{fig:fig4}(c)).

While the correlation analysis should, in principle, contain complete information about all of the noise in our experiments, its implementation is expected to become more computationally costly when applied to codes at larger distances. We simplified the decoder graph, and thus the number and size of hyperedges needed in the correlation analysis, by feeding-forward information from each round of flag measurements. This procedure, known as ``deflagging'' (\hyperref[sec:deflagging]{Supp.~\nameref{sec:deflagging}}), allowed us to eliminate all 30 of the size-4 hyperedges in an experiment with 10 rounds of stabilizer measurements without a significant increase in the logical error per round (Fig.~\ref{fig:fig4}). Furthermore, the logical errors were mostly preserved compared to results without the deflagging procedure.

Nai\"vely extending the HH code to distance-3 would result in size-5 hyperedges arising in the decoder hypergraph. However, when deflagging is applied, we found that there were no longer any size-5 hyperedges, and the number of size-4 hyperedges reduced from $148r-12$ to $60r-12$, where $r\ge1$ is the number of rounds. Since the computational resources scale exponentially with the largest weight hyperedge in a graph, we expect that the deflagging procedure will provide a dramatic reduction in the computational resources needed to carry out the correlation analysis for codes beyond distance-3.

\section{\label{sec:conclusion}Conclusion}
Experimentally preparing and repeatedly stabilizing a logical quantum state, with error rates nearly ten times smaller than the lowest physical measurement error rate, is an important step towards executing larger, fault-tolerant circuits. The hexagonal lattice on which we demonstrated our findings can be extended to operate larger distance versions of the fault-tolerant HH code used here, or for other related codes~\cite{hastingsDynamicallyGeneratedLogical2021,gidneyFaultTolerantHoneycombMemory2021,woottonHexagonalMatchingCodes2021}. Although the distance-2 version was implemented on a subset of qubits within a hexagonal lattice, other topologies are also expected to benefit; for example, a heavy-square topology akin to the rotated surface code with added flag qubits~\cite{chamberlandTopologicalSubsystemCodes2020}. Nevertheless, our probabilistic error correction methods and higher order error correlation analysis represents an approach for improving decoders for codes with or without flags within any device topology. We also demonstrated an effective use of flags to limit the extent of the correlations needed for efficient decoding. Our approach for extracting quantitative noise figures from the experiments creates a path to diagnose and reduce the logical errors per round of codes at larger distances.

As quantum computing devices become larger and less noisy, approaches such as ours may form the basis for efficiently decoding experimentally relevant errors. Other decoding strategies such as maximum-likelihood algorithms are known to scale unfavorably with code distances but may also benefit from our approach~\cite{bravyiEfficientAlgorithmsMaximum2014,bravyiDoubledColorCodes2015,heimOptimalCircuitLevelDecoding2016}. Eventually, decoders will need to be trained in real-time~\cite{dasLILLIPUTLightweightLowLatency2021}, whereby logical operations could be interleaved with calibration circuits to periodically update the decoder graph's prior information with calibrated correlation probabilities. Previously studied bootstrapping techniques~\cite{googlequantumaiExponentialSuppressionBit2021} coupled with the periodic re-calibration of the correlation edges may eventually approach near-optimal decoding efficiencies, although the existence of an optimal strategy remains an open question.

\bibliography{references_2}

\begin{acknowledgments}

We wish to acknowledge Ben W. Reichardt (USC) for proposing the deflagging procedure, Joe Latone for discussions on optimizing the numerical analysis, Ted Thorbeck for initial suggestions on readout tune-up, David Lokken-Toyli and Oliver Dial for discussions on characterizing measurement impact on qubit states, and Isaac Lauer, Andrew Eddins, David McKay and Sarah Sheldon for valuable discussions. These results were enabled by the work of the IBM Quantum software and hardware teams. We acknowledge support by IARPA under contract W911NF-16-1-0114 for the theoretical and experimental work (including partial device bring-up, characterization, and gate and measurement calibration) presented in this manuscript. The device was designed and fabricated internally at IBM. 

E.H.C. and T.J.Y. contributed equally to this work. M.L. and A.W.C. performed simulations and analysis. Y.K., N.S., S.S., A.D.C. and M.T. designed and conducted the experiments. All authors contributed to writing the manuscript.

\end{acknowledgments}

\clearpage
\newpage

\beginsupplement

\section*{Supplementary Material}
\subsection{Code preparation and measurement}
\labelname{A}\label{sec:code}
Here we describe briefly how to prepare, maintain, and measure logical states of the $\llbracket4,1,2\rrbracket$ code. A set of stabilizer generators for this code is
\begin{equation}\label{eq:stabilizers}
\mathcal{S}=\langle S_Z^{(0,2)}=ZIZI, S_X=XXXX, S_Z^{(1,3)}=IZIZ\rangle
\end{equation}
and its logical operators can be chosen as $X_L=XIXI$ and $Z_L=ZZII$, where the ordering of the Pauli operators on data qubits `0' through `3' are indexed from left to right. The stabilizer formalism implies logical states
\begin{align}\label{eq:logical_states}
\ket{0}_L&=(\ket{0000}+\ket{1111})/\sqrt{2},\\\nonumber
\ket{1}_L&=(\ket{1010}+\ket{0101})/\sqrt{2},
\end{align}
which are $\pm1$-eigenstates of $Z_L$, respectively. The $\pm1$-eigenstates of $X_L$ are $\ket{+}_L=(\ket{0}_L+\ket{1}_L)/\sqrt{2}$ and $\ket{-}_L=(\ket{0}_L-\ket{1}_L)/\sqrt{2}$.

Rather than create these logical states via unitary evolution, which may spread errors in a non-fault-tolerant fashion, instead we use projective measurements. For example, the $\ket{0}_L$ state is prepared beginning with the state $\ket{0000}$, which is already a $+1$-eigenstate of $S_Z^{(0,2)}$, $S_Z^{(1,3)}$, and $Z_L$, and measuring $S_X$ using the X-check circuit from Fig.~\ref{fig:layout}(c). Without circuit noise, if the measurement reports $0$ ($+1$-eigenstate), then we have prepared $\ket{0}_L$, and if it reports $1$ ($-1$-eigenstate), we have prepared $(IZII)\ket{0}_L$. In the second case, we could obtain $\ket{0}_L$ by applying $Z$ to the last qubit, but instead we just flip all subsequent measurements of $S_X$ to account for it.

We likewise prepare $\ket{1}_L$ by starting with $X_L\ket{0000}$ and measuring $S_X$, $\ket{+}_L$ by starting with $\ket{++++}$ and measuring $S_Z^{(0,2)}$ and $S_Z^{(1,3)}$, and $\ket{-}_L$ by starting with $Z_L\ket{++++}$ and measuring $S_Z^{(0,2)}$ and $S_Z^{(1,3)}$. Again, in each case, we may end up preparing a state that differs by a Pauli from the logical state we intended (there is one such case for $\ket{0}_L$ or $\ket{1}_L$ and three cases for $\ket{+}_L$ or $\ket{-}_L$), but this is accounted for by subsequently flipping the appropriate stabilizer measurements in the rest of the circuit.

After state preparation the circuit continues by repeated measurements of the stabilizers, alternating X- and Z-check circuits from Fig.~\ref{fig:layout}. If the state preparation required measuring $S_Z^{(0,2)}$ and $S_Z^{(1,3)}$ (the $\ket{+}_L$ and $\ket{-}_L$ cases), then a ``round" of syndrome measurement consists of the X-check circuit followed by the Z-check circuit. Alternatively, if the state preparation required measuring $S_X$ (the $\ket{0}_L$ and $\ket{1}_L$ cases), then a round of syndrome measurement consists of the Z-check circuit followed by the X-check circuit. The number of rounds $r$ is a parameter that we swept up to 10 in our experiments.

Following the $r$ rounds of syndrome measurements, all four code qubits are measured in the same basis. If we prepared states $\ket{0}_L$ or $\ket{1}_L$ they are measured in the $Z$-basis, and if we prepared $\ket{+}_L$ or $\ket{-}_L$ they are measured in the $X$-basis. From this measurement information, the values of some stabilizers and one of the two logical operators ($Z_l$ or $X_L$) can be inferred. 

See Fig.~\ref{fig:layout}(c) for an example of the circuitry described in this section for the $\ket{-}_L$ case.

\subsection{Summary of experiments}
\labelname{B}\label{sec:summary}
All experiments were taken with $120,000$ shots. In anticipation of executing larger distance codes on the heavy hexagon lattice, a variety of experimental configurations were explored:
\begin{enumerate}
    \item Code layouts \\
    $ZXZ$ : $\mathcal{S}=\langle ZIZI, XXXX, IZIZ\rangle$ and \\
    $XZX$ : $\mathcal{S}=\langle XIXI, ZZZZ, IXIX\rangle$.
    \item All 7 possible sets of 7 physical qubits needed for $\llbracket4,1,2\rrbracket$ on the lattice of 27 qubits.
    \item Initial states $|0\rangle_L$, $|1\rangle_L$, $|+\rangle_L$, $|-\rangle_L$.
    \item Logical state definitions \\
    $Z_L = ZZII$ vs. $Z_L = IIZZ$ or\\ 
    $X_L = XIXI$ vs. $X_L = IXIX$
    \item With and without dynamical decoupling during idling periods.
    \item \textit{Both} (in main text) vs. \textit{single} stabilizer measurement checks (i.e. only X-stabilizers in each round)
    \item With and without stabilizer measurement rounds of identical duration - showing that stabilizer checks preserved logical states with higher fidelity than simply idling the qubits.
\end{enumerate}

\subsection{State tomography}
Upon preparing the logical states using the projective protocol described in \hyperref[sec:code]{Supp.~\nameref{sec:code}}
we performed quantum state tomography of the four data qubits. We used the methodology implemented in Qiskit Ignis (Qiskit version \texttt{0.23.0}, Terra version \texttt{0.17.0}, Ignis version \texttt{0.6.0})~\cite{Qiskit} and ran measurement error mitigation~\cite{bravyiMitigatingMeasurementErrors2021} in its full noise matrix variant also as offered in Ignis. The reconstructed 4-qubit density matrix $\rho_{4Q}$ was then used to compute the state fidelity as $F_{4Q} = \bra{\psi}\rho_{4Q}\ket{\psi}$ where $|\psi\rangle$ is one of the logical states described in Eq.~\ref{eq:logical_states}, or the equivalent logical states in the $X-$basis. We used the \texttt{cvx} method for state reconstruction in the \texttt{StateTomographyFitter} class (see Qiskit Ignis documentation for more details). We further projected the resulting 4-qubit density matrix $\rho_{4Q}$ onto the logical codespace~\cite{takitaExperimentalDemonstrationFaultTolerant2017c,andersenRepeatedQuantumError2020a,marquesLogicalqubitOperationsErrordetecting2021} to obtain the logical codespace probability, $F_L$, along with an acceptance probability, $P_L = F_{4Q}/F_L$, as shown in Table \ref{tab:state_tomo}.

Suppose we focus on the preparation of $\ket{0}_L$ as an example. Without any errors, state tomography would give $\rho_{4Q}=(\rho_0+\rho_1)/2$, where $\rho_0$ is the state expected after measuring `0' for $S_X$ (namely, $\ket{0}_L$) and $\rho_1$ is the state expected after measuring `1' (namely, $IZII\ket{0}_L$). The state $\rho_{4Q}$ is an equal classical mixture of the two cases because they occur with equal probability. Now, define
\begin{align}
F_{4Q} &\equiv \text{tr}[\rho_{4Q}(\rho_0+\rho_1)],\\
F_L &\equiv \text{tr}[\rho_{4Q}(\rho_0+X_L\rho_0X_L+\rho_1+X_L\rho_1X_L)],
\end{align}
which represent the probabilities we have the expected logical state and the probability we are in the codespace, respectively. Note, $P_L=F_{4Q}/F_L$ is called the acceptance \textit{probability} because it is bounded between `0' and `1'.

The cases for $\ket{1}_L$, $\ket{+}_L$, and $\ket{-}_L$ are analogous. It is worth noting that in the $\ket{+}_L$ and $\ket{-}_L$ cases, $\rho_{4Q}$ is the equal mixture of four states, corresponding to the four possible combinations of measurement outcomes for $S_Z^{(0,2)}$ and $S_Z^{(1,3)}$. Also, in the $\ket{0}_L$ and $\ket{1}_L$ cases, we keep only those runs with trivial flag measurements, as non-trivial flag measurements means an error must have occurred.

For these state tomography experiments, we applied readout correction~\cite{bravyiMitigatingMeasurementErrors2021} by constructing a noisy measurement basis from the readout calibration matrix whose projectors are then used in the state reconstruction, as detailed in Ref.~\cite{garionExperimentalImplementationNonClifford2021}.

\begin{table}[h!]
	\centering
	\begin{tabular}{| c | c | c | c | }
		\hline
		logical state  & $F_{4Q}$ & $F_{L}$ & $P_L$ \\
		\hline \hline
		$\ket{0}_L$& $0.93173$ & $0.99930$ & $0.93238$\\
		\hline
		$\ket{1}_L$& $0.92532$ & $0.99950$ & $0.92580$\\
		\hline
		$\ket{+}_L$& $0.95602$ & $0.99986$ & $0.95616$\\
		\hline
		$\ket{-}_L$& $0.95632$ & $0.99988$ & $0.95644$\\
		\hline
	\end{tabular}
	\caption{\textbf{Initial state preparation state tomography fidelity using dynamical decoupling.} Physical qubit state tomography fidelity ($F_{4Q}$), logical codespace fidelity ($F_L$), and the acceptance probability ($P_L$).}
	\label{tab:state_tomo}
\end{table}

\subsection{Pauli fault tracing}
\labelname{D}\label{sec:paulitracing}
A standard model of faults in quantum error-correction is Pauli depolarizing noise: any qubit initialization, gate, idle location, or measurement can suffer a fault, in which case it is followed (or preceded, in the case of a measurement) by a Pauli $P$ acting on the same number of qubits (1 or 2) as the circuit component. Initializations and measurements can just suffer $X$ errors (as $Z$ errors have no effect), while 1- and 2-qubit gates can suffer any error from the 1- or 2-qubit Pauli groups.

\begin{figure}[h!]
\includegraphics{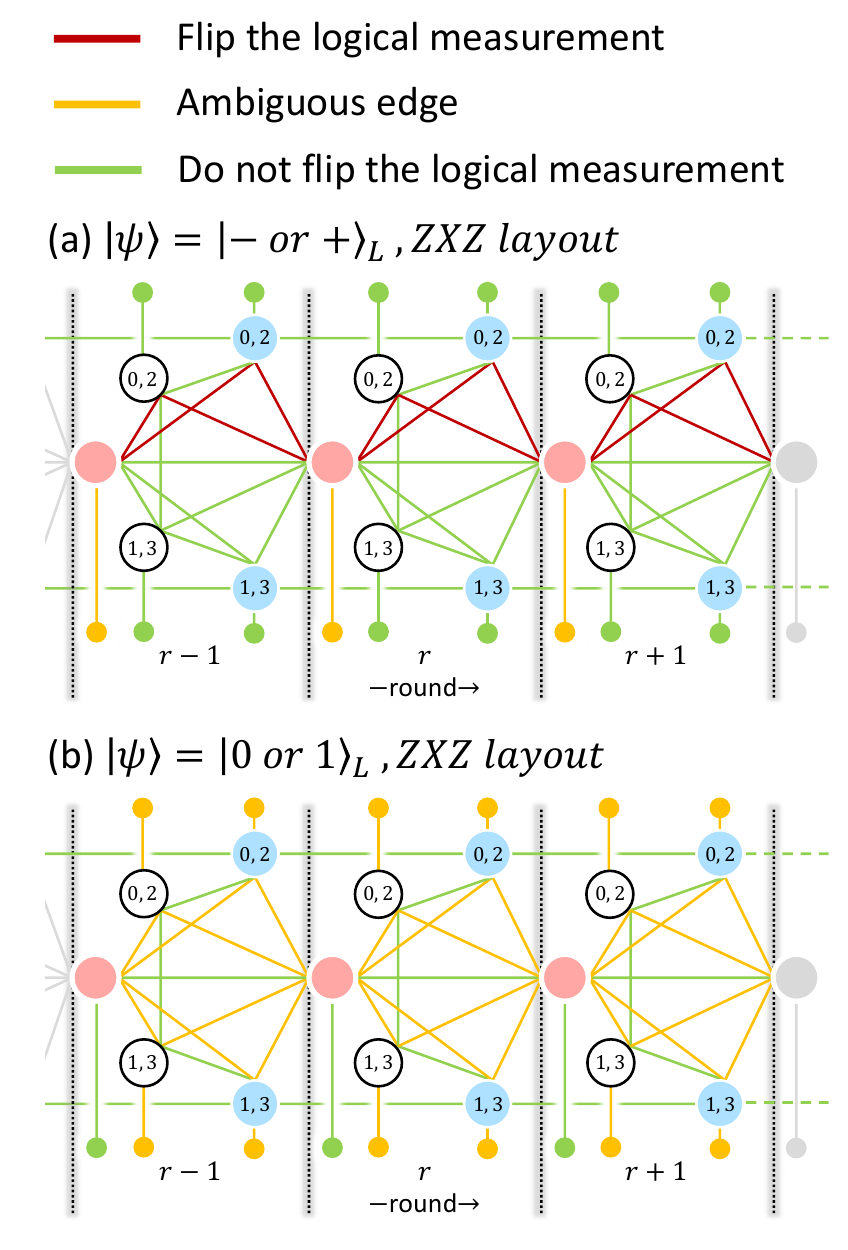}
\caption{(Color)
Three classes of edges found in the (a) $|-/+\rangle_L$ and (b) $|0/1\rangle_L$ decoder graphs for the $ZXZ$ code layout (\hyperref[sec:summary]{Supp.~\nameref{sec:summary}}). \textit{Partial} post-selection is done by excluding instances where the decoder highlighted any ambiguous (orange) edges. For $d>2$ quantum error correcting codes, there would not be any ambiguous (orange) edges in the decoding graph.
}
\label{fig:decoderEdges}
\end{figure}

Consider the set of Pauli errors that result from single faults in the syndrome measurement circuits. For each  Pauli error in this set, propagate it through the circuit and determine the set of error-sensitive events that detect the error. This set becomes a hyperedge in the decoding hypergraph. At first order, the probability $P$ of that hyperedge is just the sum of probabilities of the faults that can cause it, and its hyperedge weight is $\log((1-P)/P)$. These hyperedges that \textit{can} be predicted and categorized for partial post-selection and decoding are shown in Fig.~\ref{fig:decoderEdges}.

Not all hyperedges in the graph end up having a probability assigned. For example, in real hardware computational leakage occurs and is not accounted for by the Pauli tracer; so hyperedges not predicted by the Pauli tracer can appear in experiments such as those in Fig.~\ref{fig:fig2}. 

\subsection{Decoding examples}
\label{sec:decodingexamples}

\begin{figure}
    \centering
    \includegraphics[width=\columnwidth]{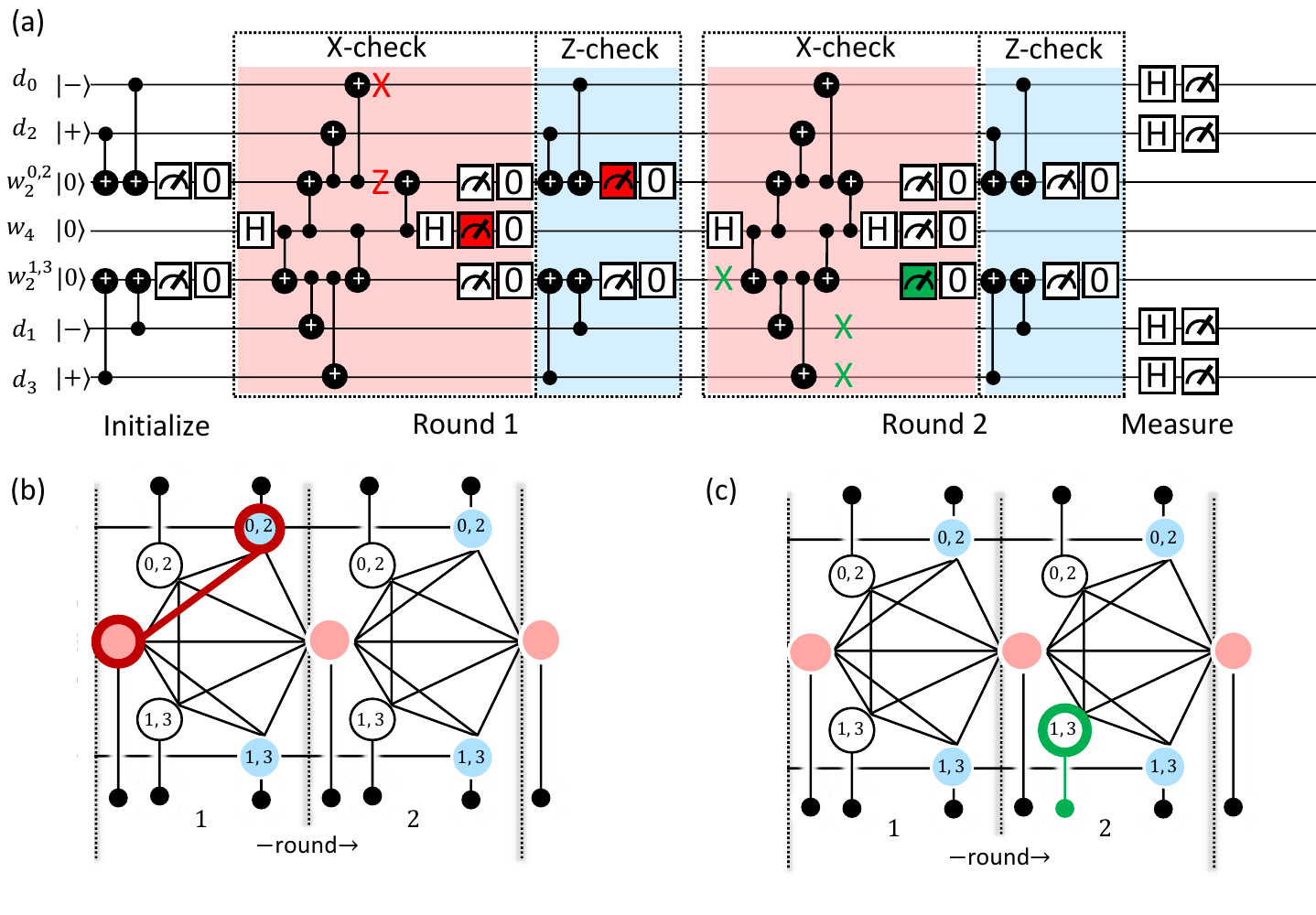}
    \caption{Single Pauli faults in a circuit (a) leading to error-sensitive events appearing in the decoder graph. (b) When a weight-2, $ZX$ Pauli error occurs after a CNOT during a X-stabilizer measurement, two events are triggered and a decoder, if tuned with the correct edge weights, would highlight the edge connecting those events. (c) When a weight-1, $X$ Pauli error occurs on a flag qubit, a weight-2 Pauli error appears on the two of the four data qubits. Without the flag measurement, this error would have gone undetected by the any subsequent X- or Z-type stabilizer measurement.
    }
    \label{fig:decodingexample}
\end{figure}

Some examples of how single Pauli faults can be used by a decoder. When certain edges are highlighted (Fig.~\ref{fig:decoderEdges}), error correction is possible for certain Pauli errors (Fig.~\ref{fig:decodingexample}(b)) while others are not (Fig.~\ref{fig:decodingexample}(c)).

\subsection{Deflagging procedure}
\labelname{F}\label{sec:deflagging}

\begin{figure}
    \centering
    \includegraphics[width=\columnwidth]{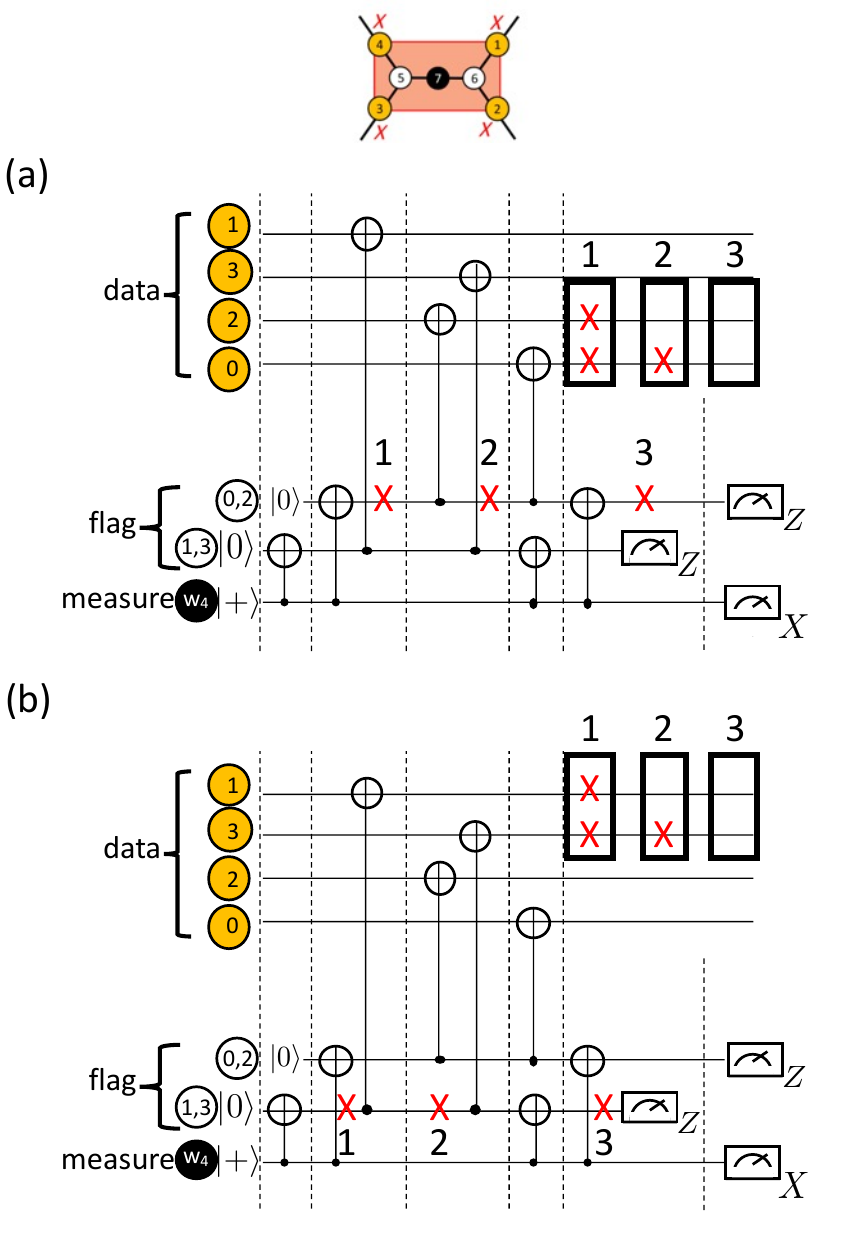}
    \caption{Deflagging procedure for $\llbracket4,1,2\rrbracket$ in a $ZXZ$ Code layout. This Pauli frame change is applied in the same manner for larger distance versions of the HH code. If both flags ($w^{0,2}$ and $w^{1,3}$) are raised, then do not apply any Pauli-$X$ in software. Otherwise:
    (a) If only the left flag ($w^{0,2}$) is raised, apply a Pauli-$X$ to data qubit ($d_0$) in software.
    (b) If only the right flag ($w^{1,3}$) is raised, apply a Pauli-$X$ to data qubit ($d_3$) in software.
    }
    \label{fig:deflagging}
\end{figure}

Using the deflagging procedure illustrated in Fig.~\ref{fig:deflagging}, the largest hyperedge sizes in the decoder hypergraph can be shown, using the Pauli tracer in \hyperref[sec:paulitracing]{Supp.~\nameref{sec:paulitracing}}, to be of size no greater than 4 for the distance-3 HH code. Generalizations of this procedure to larger distances will be discussed in upcoming work.

\subsection{Correlation analysis}
\labelname{G}\label{sec:CorrelationAnalysis}
The goal of the correlation decoder is to learn hyperedge probabilities from a set of measurement data. To do so, we assume that hyperedges occur independently. While this is not strictly true in the standard model of depolarizing noise where faults are mutually exclusive (e.g.~ if a Hadamard gate fails with an $X$ error it cannot simultaneously fail with a $Y$ error), it is easy enough to find an independent error model that is equivalent to the exclusive one (see \footnote{C. Gidney, Decorrelated depolarization, https://algassert.com/post/2001 (2020), accessed:2021-08-23.} and \cite{chaoOptimizationSurfaceCode2020}), justifying the assumption of independent hyperedges.

Denote the set of error-sensitive events by $E$ and the set of possible hyperedges by $H$. One can determine $H$ from, for instance, Pauli tracing of single faults with additional hyperedges added if they are suspected to be of experimental relevance. From measurement data, one has access to estimates of the expectation values $\langle h\rangle:=\langle\prod_{i\in h}X_i\rangle$, where $X_i$ is the binary random variable associated to error-sensitive event $i\in E$ and $h\in H$ is a hyperedge. Also, these expectation values can be written in terms of hyperedge probabilities $\alpha_h$. Suppose $L_h\subseteq H$ is the set of hyperedges that have non-empty intersection with $h$. Then we have
\begin{equation}\label{eq:expectation}
\langle h\rangle = \sum_{\substack{A\subseteq L_h\text{\space with}\\h\subseteq\triangle_{a\in A}a}}\left(\prod_{a\in A}\alpha_a\prod_{b\in L_h-A}(1-\alpha_b)\right),
\end{equation}
where $\triangle$ denotes the symmetric difference of sets. Writing these equations for all $h\in H$, one in principle has a system of $|H|$ equations and $|H|$ unknowns that can be solved for $\alpha_h$ in terms of the experimentally estimated expectations $\langle h\rangle$. In practice, this system of equations is too expensive to solve, even numerically, for $|H|\gtrsim20$. For instance, a $\llbracket4,1,2\rrbracket$ experiment preserving the logical $\ket{+}$ state for $r$ rounds of syndrome measurement has $|H|=32r-2$, which is already prohibitively large for $r=1$.

Therefore, we must settle for an approximate solution to the equations. This proceeds as follows. First, \emph{cluster} hyperedges by finding a subset $C\subset2^E$ (where $2^E$ is the powerset of $E$) such that for all $h\in H$, there is a $c\in C$ such that $h\subseteq c$. A simple approach for clustering is to sort hyperedges by size, from largest to smallest. Go through the sorted list, placing a hyperedge into $C$ if it is not already a subset of some element of $C$. We refer to elements of $C$ as \emph{clusters}. It is important for what follows that clusters are small, and it is evident from the prescribed clustering approach that the largest cluster size is equal to the largest hyperedge size.

\begin{figure}
    \centering
    \includegraphics[width=0.8\columnwidth]{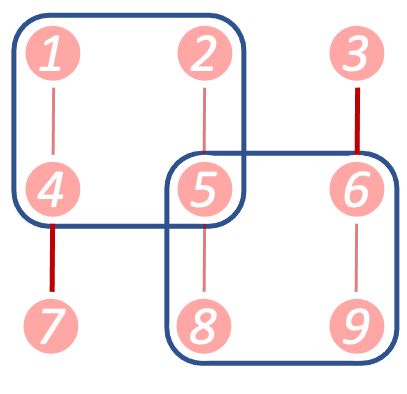}
    \caption{An example hypergraph where each node corresponds to an error-sensitive event. When an error event occurs, the node can be thought of as being highlighted. In this example, there are two size-4 hyperedges (blue squares), six size-2 hyperedges (red vertical lines), and nine size-1 hyperedges (the individual nodes). A valid set of clusters $C$ consists of both size-4 hyperedges (say $h$ and $h'$) and the two size-2 hyperedges drawn in bold, $\{3,6\}$ and $\{4,7\}$. After solving for $\hat \alpha_h$ on each cluster, hyperedges $\{4\}$, $\{5\}$, and $\{6\}$ will need their probabilities adjusted to yield the final value of $\alpha_h$.}
    \label{fig:hypergraph_example}
\end{figure}

Next, \emph{solve} each cluster $c\in C$. Suppose $S_c\subseteq H$ is the set of hyperedges that are a subset of $c$. For each $h\in S_c$, calculate $\langle h\rangle$ as if $S_c$ are the only hyperedges that exist. That is,
\begin{equation}\label{eq:cluster_expectation}
\langle h\rangle_{c} = \sum_{\substack{A\subseteq S_c\text{\space with}\\h\subseteq\triangle_{a\in A}a}}\left(\prod_{a\in A}\alpha_a\prod_{b\in S_c-A}(1-\alpha_b)\right).
\end{equation}
Now we have a system of $|S_c|$ equations and $|S_c|$ unknowns, the $\alpha_h$ for all $h\in S_c$. If clusters are small, this is fast to solve. In particular, a size-2 cluster can be solved analytically (see \cite{googlequantumaiExponentialSuppressionBit2021}), while clusters with sizes three and four can be solved numerically. In general, a cluster with size $|c|$ leads to at most $2^{|c|}-1$ equations.

The cluster solving procedure is approximate because clusters are solved assuming only hyperedges within them exist, while in actuality some hyperedges span different clusters. The final step is to adjust solutions based on these spanning hyperedges.

An example of the idea is the following (Fig.~\ref{fig:hypergraph_example}). Suppose $h\subseteq c$ is a hyperedge within cluster $c$, which has some probability $\alpha_h$. Suppose another hyperedge $h'$ exists, and $h'\not\subseteq c$ but $h'\cap c=h$. When we solved cluster $c$, we obtained some probability $\hat \alpha_h$ for hyperedge $h$, but because we ignored $h'$, $\hat \alpha_h$ is actually the sum of two different events: either $h$ occurred without $h'$ occurring, or $h'$ occurred without $h$ occurring. Therefore, $\hat \alpha_h=\alpha_h(1-\alpha_{h'})+(1-\alpha_{h})\alpha_{h'}$, and so $\alpha_h=(\hat \alpha_h-\alpha_{h'})/(1-2\alpha_{h'})$ is the probability of $h$ \emph{adjusted} by $h'$. Adjustment commutes -- if we need to adjust $h$ by several hyperedges, we can adjust by one at a time in any order.

Maximum-size hyperedges do not require adjustment, since there is no larger $h'$ to adjust by, and they provide a base case for the recursive adjustment of all smaller size hyperedges. This proceeds as follows: (1) Adjust each hyperedge $h\in H_c$ of size $s-1$ by finding all hyperedges $h'$ with weight at least $s$, such that $h'\cap c=h$ and $h'\not\subseteq c$. (2) For all such $h'$, perform the update $\alpha_h\leftarrow (\alpha_h-\alpha_{h'})/(1-2\alpha_{h'})$. After doing this for all $h'$ we are left with an adjusted $\alpha_h$. (3) Finally, because $h$ might be in several different clusters, we might have multiple adjusted $\alpha_h$. Average the adjusted values to get a final $\alpha_h$. 

We highlight a point of caution. If one executes step (1) not starting with the largest hyperedges in the graph with weight $s$, but instead only with size-2 hyperedges (which was done for the repetition code in \cite{googlequantumaiExponentialSuppressionBit2021}), then one could arise at non-physical values for $\alpha_h$ (Fig.~\ref{fig:negativePi}). 

\begin{figure}
    \centering
    \includegraphics[width=\columnwidth]{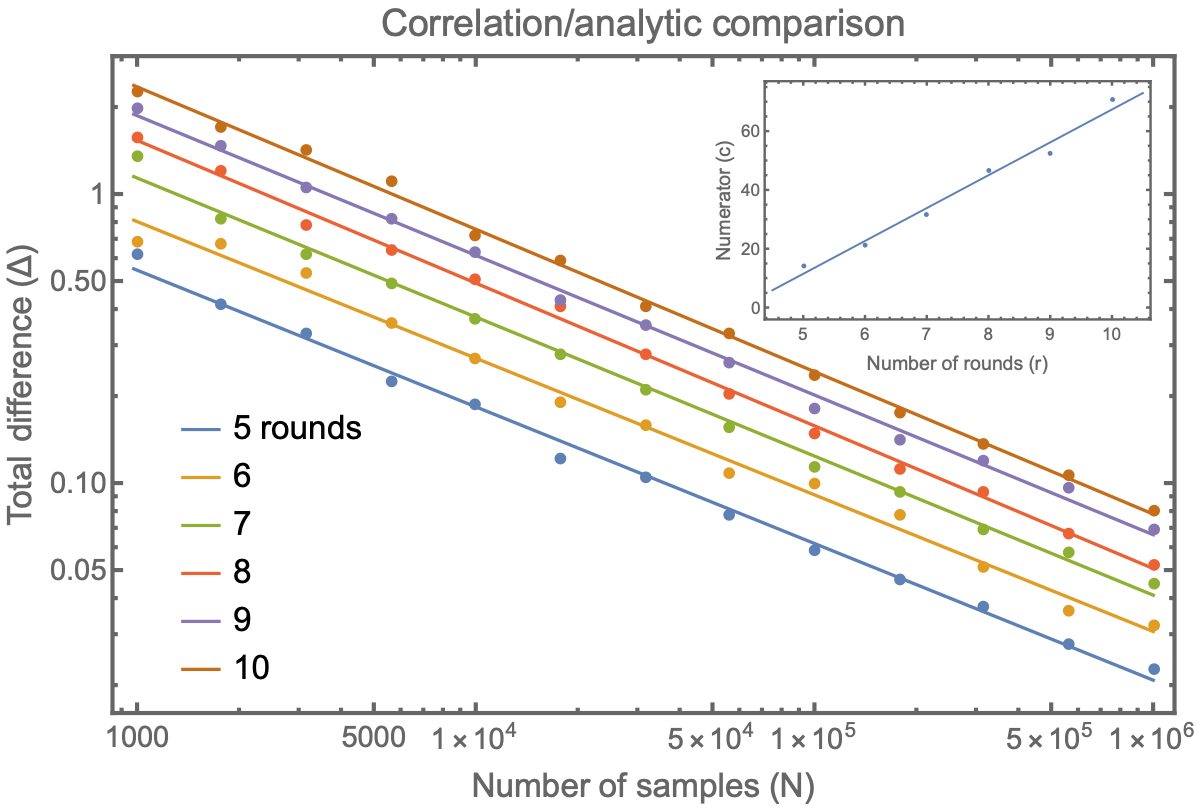}
    \caption{(Color)
    We compare the correlation decoder with the analytic decoder in simulations of $r$ rounds of the $\llbracket 4,1,2\rrbracket$ code by plotting $\Delta=\sum_{h\in H}|\alpha^{\text{correlation}}_h-\alpha^{\text{analytic}}_h|$, where $\alpha^{\text{analytic}}_h$ are hyperedge probabilities calculated at first-order in fault probabilities. If $N$ denotes the number of samples, best fits indicate the behavior $\Delta(N,r)=c(r)/\sqrt{N}$, where $c$ is a linear function.}
    \label{fig:correlation_error}
\end{figure}

\begin{figure}[h]
\includegraphics{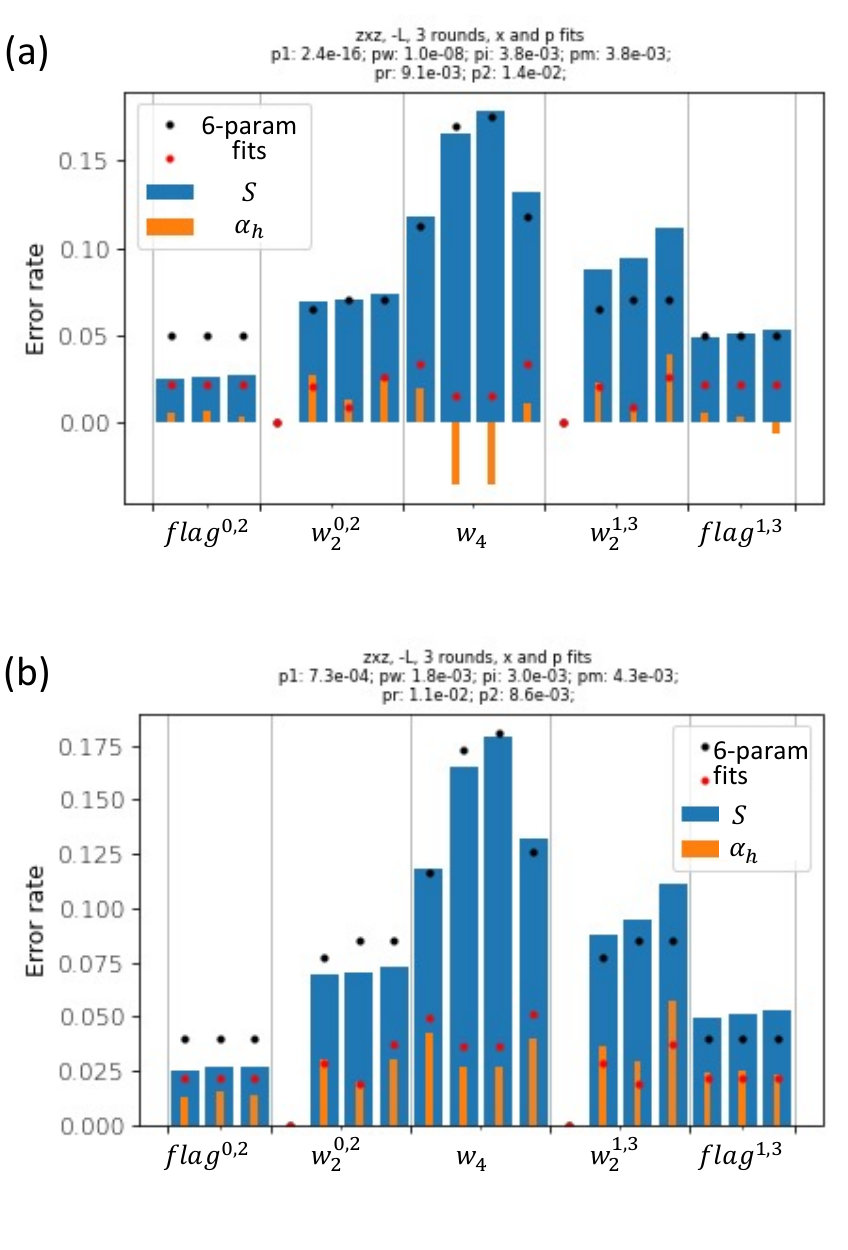}
    \caption{(Color)
    \label{fig:negativePi}
    (a) Adjustment procedure applied on the data in Fig.~\ref{fig:fig3} only up to size-2 hyperedges resulting in negative $\alpha_h$ values which are non-physical.
    (b) Adjustment procedure applied up to size-4 hyperedges resulting in size-1 values being non-negative and thereby applicable for calibrating the decoder graph edges.
    }
    
\end{figure}

In Fig.~\ref{fig:correlation_error}, we provide simulations that suggest the correlation analysis is providing an accurate assessment of the hyperedge probabilities. The error in the correlation analysis scales with the number of runs of the experiment, $N$, as $1/\sqrt{N}$. 

\subsection{\textit{Partial} post-selection}
\label{sec:partialPS}

To implement \textit{partial} post-selection with the $\llbracket4,1,2\rrbracket$ code, the edges highlighted by the MWPM decoding algorithm need to be compared against the three classes of edges classified in Fig.~\ref{fig:decoderEdges}. The illustrated decoder graph will change depending on the logical state and the code layout. Also, the ambiguous edges do not appear for $d>2$ codes in cases with only single faults per round, and so such a \textit{partial} post-selection scheme is no longer applicable in those cases.

\subsection{\label{sec:methods:device}Device properties}

Characterized noise properties of the 7 qubits used in \textit{ibmq\_kolkata} are shown in Table~\ref{table:deviceRB}. All 1-qubit gates had a duration of 35.55~ns. The measurement pulse width and integration windows were approximately 330~ns. The total measurement and conditional reset cycling time, including delays from the cable transmission and electronic latency, were approximately 764~ns. Table~\ref{table:deviceCX} shows the result of optimizing the error per gate for each entangling, cross-resonance rotary echo gate~\cite{sundaresanReducingUnitarySpectator2020}. The direction of the cross-resonance gates were chosen to minimize the impact of spectator qubits. Composing these operations resulted in Z-(X-)stabilizer checks lasting for $1.66 \mu s$ ($2.93 \mu s$).

\subsection{Calibration of measurement power}
\label{sec:measurementcalibration}
\begin{figure}[h]
\centerline{\includegraphics[width=1\columnwidth]{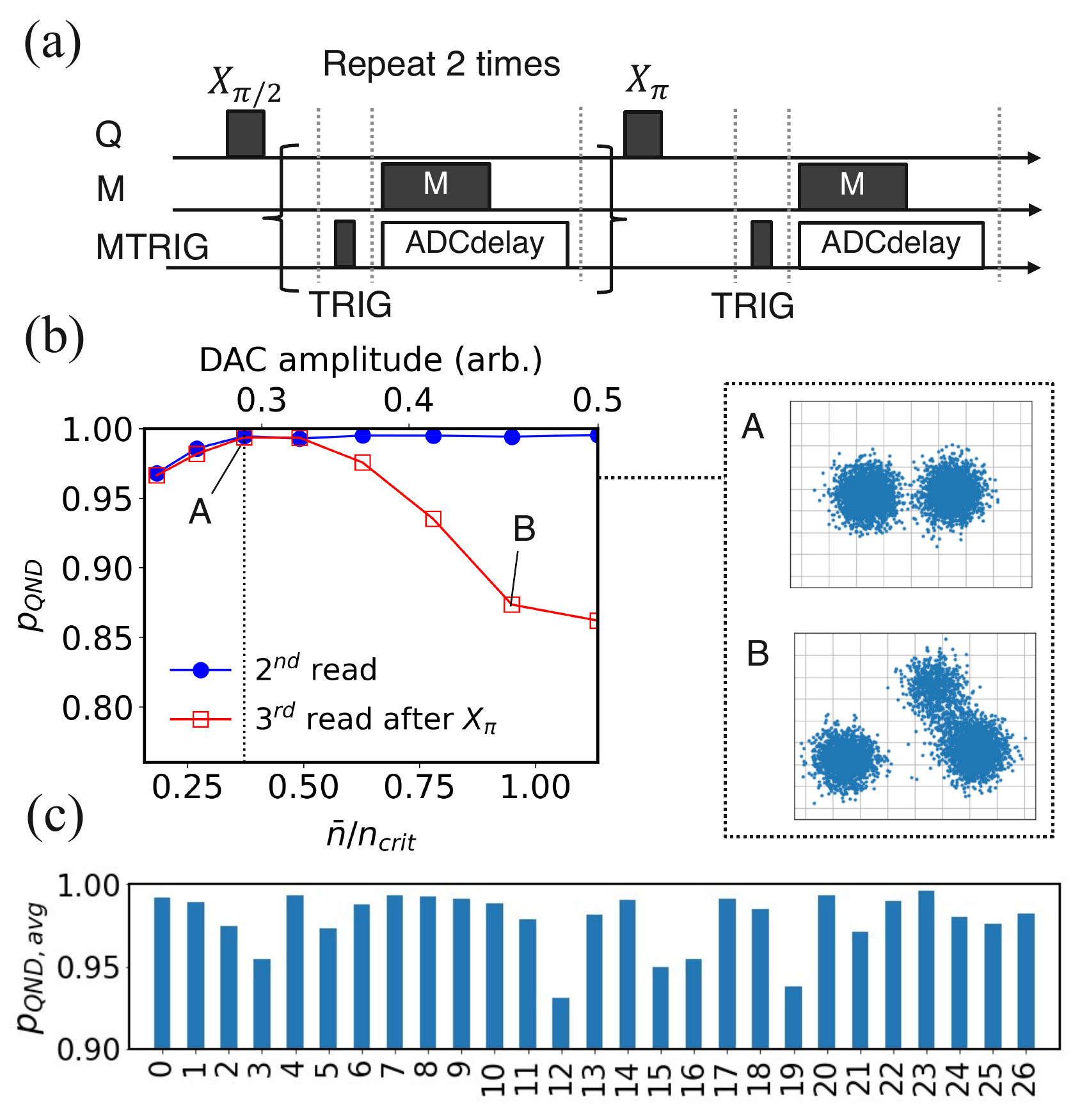}}
\caption{(Color)
\label{fig:meas_tuneup} 
(a) Illustration of a readout measurement protocol for optimizing the mid-circuit measurement amplitudes with a pulse schedule for the qubit drive (Q), measurement drive (M), and trigger for measurement (MTRIG) channels. (b) A representative QND measurement sweep for a single qubit (Q17) using the repeated readout schedule in (a). The I-Q plane scatter plot is shown as an inset for two points along the curve, where two clusters in `A' represent the ground and excited states, and a third cluster in `B' corresponds to the second excited state. The DAC amplitude was optimized by maximizing the average $p_{QND}$ from two consecutive measurements, as indicated by a dashed vertical line. (c) The resulting $p_{QND,avg}$ for the optimized readout power across all 27 qubits. The index for each qubit is indicated on the horizontal axis. 
}
\end{figure}

An optimal mid-circuit measurement tone needs to have sufficiently high power to distinguish the $\ket{0}$ and $\ket{1}$ states without inducing substantial measurement back-action. We optimized our mid-circuit measurement tones by utilizing the repeated measurement protocol illustrated in Fig.~\ref{fig:meas_tuneup}(a). The protocol involves preparing the state in a superposition of $\ket{0}$ and $\ket{1}$ using a $\gate{X_{\pi/2}}$ pulse, and then concluding with two sequential readout pulses. The outcome of the first readout pulse should be random, while the second result should, ideally, match the first if the state was not impacted by measurement back-action or poor readout fidelity. We quantify the degree of measurement-induced back-action using the quantum non-demolition (QND) probability defined as 
\begin{equation}
    p_{QND}=[p(0 \vert 0)+p(1 \vert 1)]/2,
\end{equation}
which tracks whether the state was unchanged from the previous measurement. However, excessive readout power can also result in excitations out of the computational basis and into leaked states, which can be incorrectly categorized as $\ket{0}$ or $\ket{1}$ by a linear state discriminator. To remedy this situation, we inserted a $\gate{X_{\pi}}$ pulse which leaves non-computational states unchanged but induces a bit-flip if the state was within the computational basis. Thus, we define 
\begin{equation}
    p_{QND,X_{\pi}}=[p(0|1)+p(1|0)]/2,
\end{equation}
which quantifies the computational leakage. Note that we also incorporated a delay pulse, $\gate{ADC delay}$, whose length was chosen to be $t_{M}+10/\kappa_m$, where $t_{M}\simeq 327$~ns was the measurement tone width and $\kappa_m/2\pi=5.331$ MHz is the median $\kappa$ of the device. The additional time delay allows the cavity to depopulate and thus prevents errors when applying the subsequent $\gate{X_{\pi}}$ operation. Figure~\ref{fig:meas_tuneup}(b) shows $p_{QND}$ and $p_{QND,X_\pi}$ as a function of the average cavity photon number $\bar{n}$ normalized by the critical photon number $n_{\rm{crit}} = \delta\Delta/[4\chi(\Delta+\delta)]$~\cite{blaisCavityQuantumElectrodynamics2004} in blue and red, respectively. In the expression above, $\delta$ is the qubit anharmonicity and $\Delta = \omega_{\rm{r}} - \omega_{\rm{q}}$ is the cavity-qubit detuning. Three different DAC amplitudes were converted to average photon numbers using the protocol described in Ref.~\cite{corcolesExploitingDynamicQuantum2021d} and the fitted curve was used for extrapolation. The particular critical photon number for this qubit was $\simeq 47.3$ ( $\chi/2\pi=-1.549$ MHz, $\delta/2\pi=-340.4$ MHz, and $\Delta/2\pi=2.123$ GHz). When the cavity was populated with low photon numbers, both $p_{QND}$ and $p_{QND,X_\pi}$ curves showed monotonically increasing trends as the state distinguishability improved and, consequently, the readout error decreased. When we reached substantially high measurement powers, we observed a gradual degradation in $p_{QND,X_\pi}$ compared with $p_{QND}$ due to the population of the $\ket{2}$ state. Note that $\ket{2}$ was incorrectly recorded as $\ket{1}$ and is not captured in the $p_{QND}$ metric. However, applying a $\gate{X_\pi}$ gate captures the transition to $\ket{2}$ as an additional degradation in $p_{QND,X_\pi}$. The optimal photon number, or DAC amplitude, was thus chosen to maximize $p_{QND,avg}=(p_{QND}+p_{QND,X_\pi})/2$. The same procedure was repeated for all qubits as illustrated in Fig.~\ref{fig:meas_tuneup}(c) resulting in an overall average value for $p_{QND,avg}\simeq 0.974\pm0.020$. 

\subsection{Non-local effect of measurements}
\subsubsection{Measurement-induced phase rotations}
The multiple readout cavities were designed specifically to have frequencies and other operating parameters to allow for simultaneous readout from the same output line. However, due to variations in the fabrication process, cavity frequencies were, in this case, closer than intended. This resulted in phase rotations on data qubits induced by unintended coupling to nearby readout cavities. While we found no evidence of dephasing detected in two qubit pairs within the same readout multiplexed line, we observed coherent phase rotations between qubits on pairs of cavities whose readout frequency separations were smaller than intended. 

\begin{figure*}[h]
\centerline{\includegraphics[width=2\columnwidth]{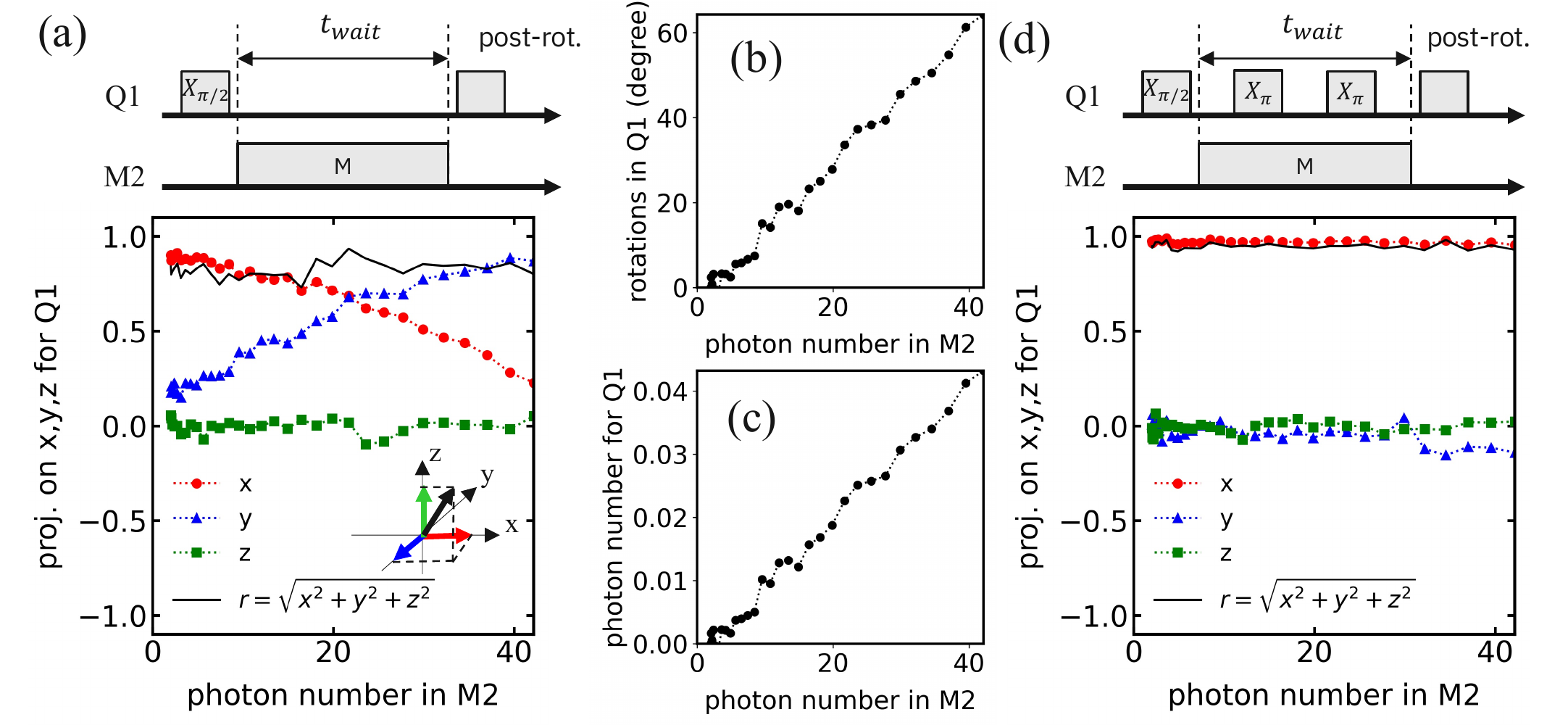}}
\caption{(Color)
(a) The pulse sequence used to monitor undesirable dephasing or rotations on Q1 while a probe measurement tone was supplied to Q2 (denoted as M2). The $x$, $y$, $z$ projections were measured using $X_{\pi/2}$, $Y_{-\pi/2}$, $I$ post-rotation gates applied, respectively, immediately before the final measurement on Q1. The photon numbers were calibrated using three different DAC amplitude using photon time operation experiment\cite{corcolesExploitingDynamicQuantum2021d}, and the DAC amplitude was converted to the corresponding photon number (denoted as photon number in M2).
The corresponding rotation angles were computed from $x$ and $y$ projections in (b). The induced rotation was assumed to be induced by a frequency shift in Q1, and corresponding photon number for Q1 is estimated in (c).
(d) The undesirable rotation was suppressed by introducing a refocusing sequence composed of $X_{\pi}-X_{\pi}$. This characterization was carried out on a separate device, and not on \textit{ibmq\_kolkata}.}
\label{fig:meas_rot}
\end{figure*}

As an egregious example of this effect, we chose a pair of qubits in a separate device whose readout frequencies turned up in practice closer than intended in design. This pair of qubits were coupled to readout cavities both coupled to the same, multiplexed line with frequencies $\sim 34$~MHz apart and therefore closer than intended. Fig.~\ref{fig:meas_rot}(a) shows an experiment for two qubits not on \textit{ibmq\_kolkata} where one qubit (Q1) is prepared with $\gate{X_{\pi/2}}$, and measured in all different axes projections, $X, Y, Z$, after a fixed delay time $t_{\rm{wait}}=1.5\mu s$. While monitoring Q1, we applied a probe measurement tone on another qubit, Q2 (M2). While we increased the probe measurement tone (M2), we observed coherent rotations on Q1, as seen on $X$- and $Y$- projections. Note that the state vector length (black line) was relatively constant, indicating that there was no significant dephasing. We computed the rotation angles from $X$- and $Y$- projections in Fig.~\ref{fig:meas_rot}(b) and inferred the photon number of the Q1 cavity by using an independently measured $\chi_1/2\pi=-1.357 \pm 0.006$ MHz. Fig.~\ref{fig:meas_rot}(c) shows that a fractional photon number was populated in the Q1 cavity, which caused an undesirable $Z$-rotation. This possible photon leakage may have induced a phase error on data qubits during syndrome measurements. Fortunately, this undesirable $Z$-rotation was corrected by inserting refocusing pulses. Fig.~\ref{fig:meas_rot}(d) illustrates the same experiment as (a) but with a dynamical decoupling sequence ($X_{\pi}-X_{\pi}$) inserted during the idling time. As a result, the measurement-induced $Z$-rotation was no longer observed. These experiments suggest that the improved logical error rates when dynamical decoupling is used can be explained by the suppression of undesirable, measurement-induced phase errors during the syndrome measurements.

\subsubsection{Measurement-induced collisions}
Our readout resonators were designed to operate at $\approx7$~GHz so that they operated above the qubit frequencies at $\approx5$~GHz. In this frequency configuration, an applied measurement tone Stark-shifts the connected qubits lower in frequency. The qubits can unintentionally be Stark-shifted into resonance with an adjacent qubit causing measurement-induced collisions, which degrade the readout fidelity.

\begin{figure}[h]
\centerline{\includegraphics[width=\columnwidth]{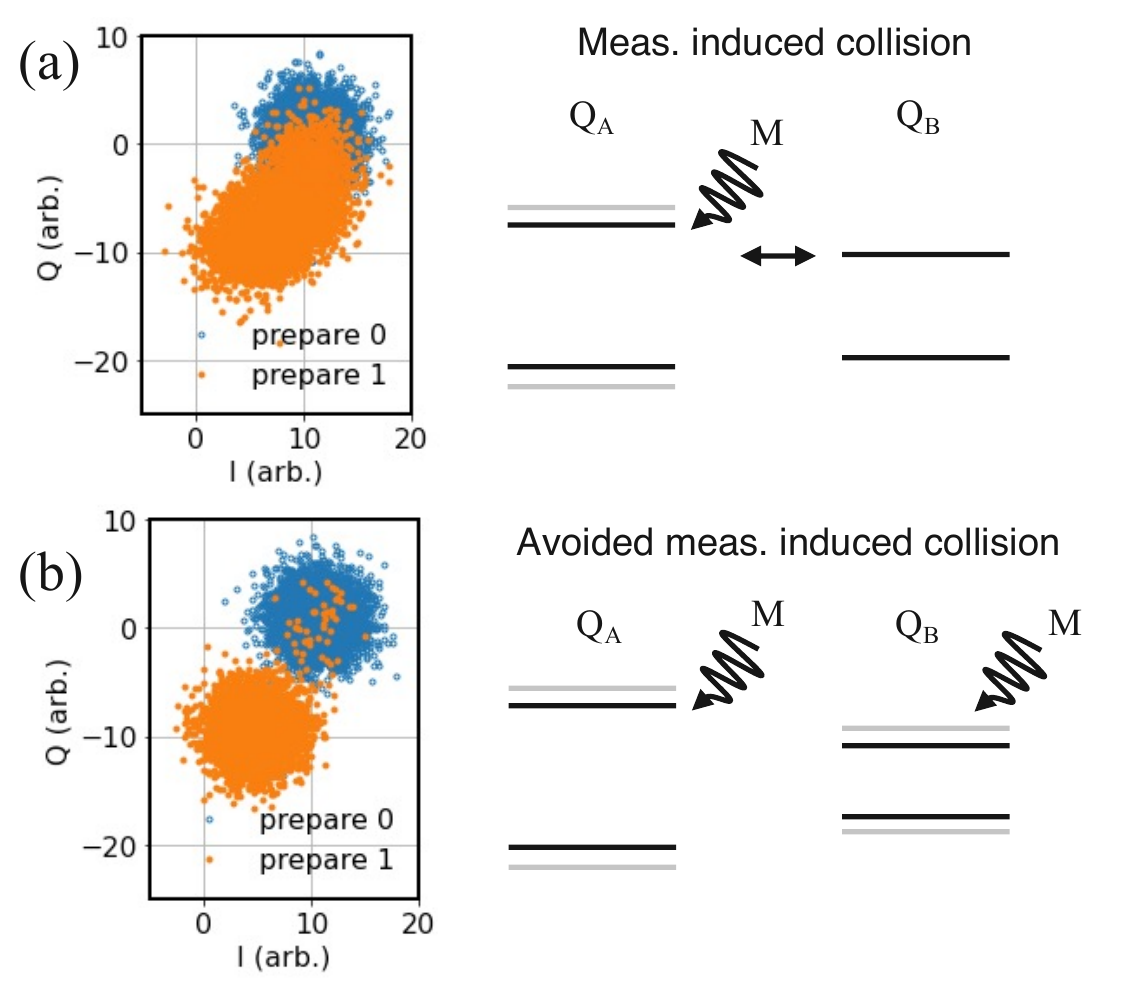}}
\caption{(Color)
I-Q scatter plots after readout of qubit $Q_A$ with 5,000 samples when the qubit was prepared in the ground (0, blue) and excited (1, orange) states. (a) without and (b) with a simultaneously applied measurement (M) tone on a neighboring qubit $Q_B$. 
The light-gray lines illustrate the effect of the Stark-shift due to the applied measurement tones.
This characterization was carried out on a separate device, and not on 
This characterization was carried out on a separate device, and not on \textit{ibmq\_kolkata}.
}
\label{fig:meas_col}
\end{figure}

Although such an effect was not isolated on \textit{ibmq\_kolkata}, we observed this effect on a pair of qubits with nearby frequencies to observe this measurement-induced collision. Fig.~\ref{fig:meas_col}(a) shows the readout scatters and energy levels of the two qubits, Q$_A$ and Q$_B$, with frequencies $\omega_{01}^{Q_A}/2\pi=4.959$ GHz and $\omega_{01}^{Q_B}/2\pi=4.921$ GHz. When a measurement tone was applied to Q$_A$, we estimated the photon number in the readout cavity to be $\bar{n}\simeq 7.738$. The corresponding Stark shift in Q$_A$'s frequency was estimated as $\delta \omega_{01}^{Q_A}/2\pi\simeq  2(\chi_A/2\pi)\bar{n}\simeq - 22.054$ MHz, resulting in a frequency closer to $\omega_{01}^{Q_B}$. The resulting frequency collision was evident in the readout scattering plot of Q$_A$ in Fig.~\ref{fig:meas_col}(a).
This measurement induced collision can be avoided by simultaneously applying another measurement tone to the adjacent qubit Q$_B$. Fig.~\ref{fig:meas_col}(b) shows the simultaneous readout where $\delta \omega_{01}^{Q_B}\simeq - 43.939$~MHz mitigated the shift $\delta \omega_{01}^{Q_A}$. This yielded the readout scattering plot in Fig.~\ref{fig:meas_col}(b), which looked much more Gaussian. The effect of measurement-induced collisions may be detrimental for readout performance of any qubit whose frequency is too close to those of its neighboring qubits, and is an important consideration for maintaining high-quality readout across many fixed-frequency qubits~\cite{hertzbergLaserannealingJosephsonJunctions2021}.

\subsection{Logical Errors With(out) Dynamical Decoupling}
\begin{figure}
\includegraphics{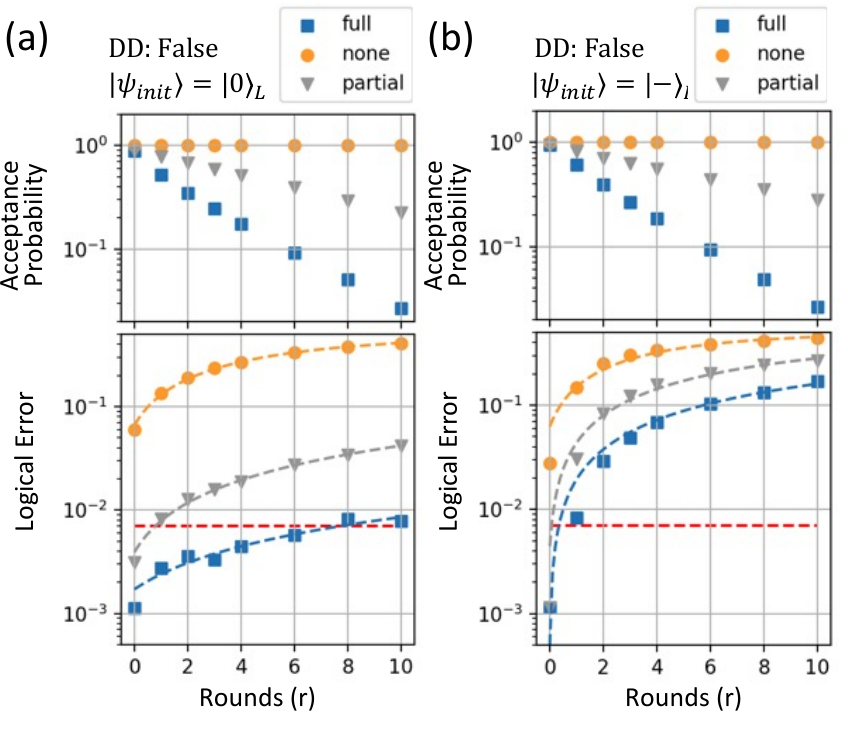}
\caption{(Color)
Identical experiment and analysis as Fig.~3 in the main text, except no dynamical decoupling (DD) sequences were employed.
(top) Fraction of total results used for the logical states ((a) $|0\rangle_L$, (b) $|-\rangle_L$) as the number of stabilizer rounds are repeated from 0 to 10 times when full (blue squares), none (yellow circles), or partial (gray triangles) partial post-selection analysis was used. 
(bottom) The corresponding logical error versus number of rounds. The dashed red lines indicate the pseudo-threshold as determined by the best (average) physical measurement errors of $7\times10^{-3}$ ($7.7\times10^{-3}$) seen in Table~\ref{table:deviceRB}.
}
\label{fig:fig3_supp}
\end{figure}

The following equation was used to fit the logical errors with (Fig.~\ref{fig:fig3}) and without dynamical decoupling (Fig.~\ref{fig:fig3_supp}):
\begin{equation}
    P_{\text{fail}}(r) = \frac{1}{2} - \frac{1}{2}(1-2\epsilon_i)(1-2\epsilon_m)e^{-2\epsilon r}
\label{eqn:logicalerr}
\end{equation}
where $\epsilon_i$ is the logical initialization error, $\epsilon_m$ the final logical data measurement error, and $\epsilon$ the logical error rate per stabilizer round (Table~\ref{table:logicalfits}).
\begin{table}
\begin{tabular}{cccccc}
\toprule
Logical &  &  &      &       &       \\
State & DD & Post-selection & $\epsilon_i$      & $\epsilon_m$      & $\epsilon$      \\ \hline
$|0\rangle_L$           & TRUE     & full     & 7.00E-04 & 1.18E-03 & 1.18E-03 \\
$|0\rangle_L$           & TRUE     & none     & 3.68E-02 & 1.04E-01 & 1.04E-01 \\
$|0\rangle_L$           & TRUE     & partial  & 2.17E-03 & 6.67E-03 & 6.67E-03 \\
$|-\rangle_L$          & TRUE     & full     & 2.83E-04 & 5.06E-04 & 5.06E-04 \\
$|-\rangle_L$          & TRUE     & none     & 1.31E-02 & 3.95E-02 & 3.95E-02 \\
$|-\rangle_L$          & TRUE     & partial  & 8.17E-04 & 9.99E-03 & 9.99E-03 \\
$|0\rangle_L$           & FALSE    & full     & 8.58E-04 & 6.90E-04 & 6.90E-04 \\
$|0\rangle_L$           & FALSE    & none     & 3.41E-02 & 7.97E-02 & 7.97E-02 \\
$|0\rangle_L$           & FALSE    & partial  & 1.97E-03 & 3.98E-03 & 3.98E-03 \\
$|-\rangle_L$          & FALSE    & full     & N/A & 1.94E-02 & 1.94E-02 \\
$|-\rangle_L$          & FALSE    & none     & 2.35E-02 & 1.25E-01 & 1.25E-01 \\
$|-\rangle_L$          & FALSE    & partial  & 2.23E-03 & 4.12E-02 & 4.12E-02 \\
\bottomrule
\end{tabular}
\caption{Fit results from fitting Eqn.~\ref{eqn:logicalerr}. Results with and without dynamical decoupling (DD) are shown. No decoding was done on data with \textit{no} post-selection.}
\label{table:logicalfits}
\end{table}

\clearpage
\newpage

\onecolumngrid

\begin{table}[h]
\begin{tabular}{@{}c | ccccccc@{}}
\toprule
                                         & $Q_0$       & $Q_1$       & $Q_2$       & $Q_4$       & $Q_6$       & $Q_7$       & $Q_{10}$   \\
                                         \midrule
         \hline
{[}{[}4,1,2{]}{]} label                  & $d_0$       & $w_2^{0,2}$ & $d_2$       & $w_4$       & $d_1$       & $w_2^{1,3}$ & $d_3$       \\ \midrule
$T_1 (\mu s)$ ($p_w$)      & 97.5     & 193.1    & 116.5    & 95.8     & 123      & 143.2    & 102.3    \\ \midrule
$T_2 (\mu s)$  & 112.4    & 217.8    & 159.9    & 34.1     & 118.4    & 21.1     & 114.7    \\ \midrule
Readout Fidelity ($1-p_m$) & 0.9910  & 0.9930  & 0.9930  & 0.9910  & 0.9930  & 0.9930  & 0.9920  \\ \midrule
1Q Error per Clifford, RB (isolated)     & 1.49E-04 & 1.42E-04 & 1.69E-04 & 1.76E-04 & 3.81E-04 & 1.46E-04 & 1.56E-04 \\ \midrule
1Q Error per Clifford, RB (simultaneous, $p_1$) & 1.99E-04 & 2.01E-04 & 1.96E-04 & 2.24E-04 & 2.91E-04 & 2.25E-04 & 2.14E-04 \\ \midrule
Conditional Reset Error, applied 1x ($p_r$)           & 0.0173   & 0.0094   & 0.0097   & 0.0066   & 0.0065   & 0.0084   & 0.0131   \\ \midrule
Conditional Reset Error, applied 2x            & 0.0156   & 0.0038   & 0.0075   & 0.0041   & 0.004    & 0.0059   & 0.0078   \\ \midrule
Conditional Reset Error, applied 3x            & 0.0172   & 0.0036   & 0.0076   & 0.005    & 0.0046   & 0.006    & 0.0077   \\ \midrule
Conditional Reset Error, applied 4x            & 0.0189   & 0.0041   & 0.0077   & 0.0045   & 0.0055   & 0.0063   & 0.0074   \\ \midrule
Unconditional Reset Error ($p_i$)                  & 0.0104   & 0.0048   & 0.006    & 0.0075   & 0.0069   & 0.0076   & 0.0077   \\ 
\bottomrule
\end{tabular}
\caption{Comprehensive characterization results of the 7 qubits used for the $\llbracket4,1,2\rrbracket$ experiments. The conditional resets were implemented using an FPGA-based conditional reset, in contrast to the unconditional reset. From this table, five of the six noise parameters were used for calculating the edge weights in the \textit{analytical} decoder graph used in Fig.~\ref{fig:fig2}. The reset error $p_r$ is defined in the supplementary material of \cite{corcolesExploitingDynamicQuantum2021d}.}
\label{table:deviceRB}
\end{table}

\twocolumngrid

\begin{table}[]
\begin{tabular}{c|c|cc}
\hline
\toprule
2-qubit Gate    & EPC        & EPC                & Simultaneous       \\ 
Control, Target & (Isolated) & (Simultaneous)     & Pair      \\ \hline
0, 1            & 0.0045        & 0.0054             & 7, 4  \\ 
2, 1            & 0.0049        & 0.0056             & 7, 4  \\ 
4, 1            & 0.0049        & 0.0059             & 10, 7 \\ 
4, 1            &                & 0.0110             & 7, 6  \\ 
7, 4            & 0.0106        & 0.0120             & 0, 1  \\ 
7, 4            &                & 0.0124             & 2, 1  \\ 
7, 6            & 0.0094        & 0.0150             & 4, 1  \\ 
10, 7           & 0.0057        & 0.0065             & 4, 1  \\ \hline \hline
Arithmetic mean & 0.0067         & 0.0092             &       \\ 
Geometric mean  & 0.0063         & 0.0085             &       \\
\bottomrule
\end{tabular}
\caption{2-qubit Errors per Clifford (EPC) from randomized benchmarking characterization done in an \textit{isolated} and \textit{simultaneous} manner. For \textit{simultaneous} benchmarking, the pair of qubits being simultaneously benchmarked is also listed. The arithmetic mean of the EPC estimated from \textit{simultaneous} benchmarking was substituted, after a slight correction explained in Table.~\ref{table:noiseSix}, for $p_2$ needed for populating the edge weights in the \textit{analytical} decoder graph used in Fig.~\ref{fig:fig2}.}
\label{table:deviceCX}
\end{table}

\begin{table}[]

\begin{tabular}{cccc}
\toprule
         & Error per Gate  &  & Randomized Benchmarking  \\ 
Variable & Description       & Fit      & (Simultaneous) \\ \hline
$p_1$       & 1-qubit           & 7.30E-04               & 2.20E-04                           \\
$p_w$       & Idle              & 1.80E-03              & 6.00E-03                           \\
$p_i$       & Initialization    & 3.00E-03             & 7.00E-03                           \\
$p_m$       & Measurement       & 4.30E-03            & 7.70E-03                           \\
$p_r$       & Readout           & 1.10E-02              & 1.00E-02                           \\
$p_2$       & 2-qubit           & 8.60E-03              & 9.00E-03                           \\
\bottomrule             
\end{tabular}
\caption{Six-parameter noise model fitting the Error per Gate (EPG) used throughout the main text. The ratio between EPC, as given in Tables \ref{table:deviceRB} and \ref{table:deviceCX}, to EPG is given by $\frac{2^n-1}{2^n}$, where $n$ is the number of qubits involved in the gate.}
\label{table:noiseSix}
\end{table}

\clearpage
\newpage

\end{document}